\documentclass{elsart}

\usepackage{epsfig,amssymb}

\def\ybco{YBa$_2$Cu$_3$O$_{7-\delta}$}

\def\tbcco{Tl$_2$Ba$_2$Ca$_2$Cu$_3$O$_{10}$}

\def\biscco{Bi$_2$Sr$_2$Ca$_2$Cu$_3$O$_{10}$}
\def\bisccoIIz{Bi$_2$Sr$_2$CaCu$_2$O$_{z}$}
\def\etal{{\it et~al.}}
\def\tJPR{t-JPR}
\def\K{\,{\rm K}}
\def\cm{\,{\rm cm^{-1}}}

\def\zo{(\omega)}
\def\e{\epsilon}

\def\eo{\epsilon_0}
\def\t{}
\def\ph{^{\rm ph}}

\def\loc{^{\rm loc}}
\def\inbl{_{\rm bl}}
\def\out{_{\rm int}}
\def\cp{_{\rm ocp}}
\def\bl{_{\rm bl}}
\def\intt{_{\rm int}}

\def\d{{\rm d}}
\def\ii{{\rm i}}
\def\ei{\epsilon_\infty}

\begin{document}

\def\o{\omega}

\begin{frontmatter}
\boldmath
\title{Phonon anomalies in trilayer high-$T_{\rm c}$ cuprate
superconductors}
\unboldmath

\author{Adam Dubroka\corauthref{cor}}
\ead{dubroka@physics.muni.cz}
\corauth[cor]{fax: +420 541 211 214, tel: +420 541 129 378}
and
\author{Dominik Munzar}
\ead{munzar@physics.muni.cz}
\address{Institute of Condensed Matter Physics, Faculty of Science, Masaryk
University,\\Kotl\'a\v{r}sk\'a 2, CZ-61137 Brno, Czech republic}
\begin{abstract}
We present an extension of the model proposed recently to account for
dramatic changes below $T_{\rm c}$ (anomalies) of some $c$-axis polarized
infrared-active phonons in bilayer cuprate superconductors, that applies to
trilayer high-$T_{\rm c}$ compounds. We discuss several types of phonon
anomalies that can occur in these systems and demonstrate that our model is
capable of explaining the spectral changes occurring upon entering the
superconducting state in the trilayer compound \tbcco. The low-temperature
spectra of this compound obtained by Zetterer and coworkers display an
additional broad absorption band, similar to the one observed in underdoped
YBa$_{2}$Cu$_{3}$O$_{7-\delta}$ and Bi$_{2}$Sr$_{2}$CaCu$_{2}$O$_{8}$. In
addition, three phonon modes are strongly anomalous. We attribute the
absorption band to the transverse Josephson plasma resonance,
similar to that of the bilayer compounds. The phonon anomalies are shown to
result from a modification of the local fields induced by the formation of
the resonance. The spectral changes in \tbcco\ are compared with those
occurring in \biscco, reported recently by Boris and coworkers.
\end{abstract}

\begin{keyword}
electron-phonon interaction \sep Tl2Ba2Ca2Cu3O10 \sep Bi2Sr2Ca2Cu3O10
\sep Tl2Ba2CaCu2O8
\PACS 74.25.Gz \sep  74.25.Kc \sep 74.72.Fq
\end{keyword}
\end{frontmatter}

\section{Introduction}
Some of the phonon modes in the high-$T_{\rm c}$ cuprate superconductors
are strongly renormalized when going from the normal to the superconducting
state~\cite{Litvinchuk,Homes,Schutzman,Bernhard,Zelezny,Boris,Zetterer,Hadjiev,Limonov,Petrov}.
It is important to establish what particular type of electron-phonon
interaction is responsible for these effects (so called phonon anomalies),
and whether the anomalous phonons are ``active players" in the mechanism of
the high-$T_{\rm c}$ superconductivity~\cite{Kulic,Lanzara,Shen} or whether
they simply respond to changes of the electronic ground state that have an
independent cause. Here we focus on the anomalies of {\it c}-axis polarized
infrared-active phonons. The most pronounced of them have been observed in
the so called multilayer compounds having two or more copper-oxygen
planes per unit cell. Recently Munzar \etal~\cite{Mike} explained phonon
anomalies in underdoped \ybco\ (Y-123). The explanation is based on the
multilayer model of van der Marel and
Tsvetkov~\cite{Marel,Gruninger,Marel2}, and the assumption of a week
(Josephson) coupling between the copper-oxygen planes: it has been
assumed that the interlayer conductivities are fairly incoherent in the
normal state and dominated by the coherent superfluid contribution in the
superconducting state. A bilayer superconductor is thus considered as a
superlattice of inter-bilayer and intra-bilayer Josephson junctions ---
this picture is called the Josephson superlattice model. The presence of
two different junctions in a unit cell gives rise to a new transverse mode,
the so called transverse Josephson plasma resonance (\tJPR) with a frequency
between the one of the inter-bilayer (longitudinal) and the intra-bilayer
(longitudinal) plasmon. The same picture of the {\it c}-axis charge
dynamics of a bilayer system has also been arrived at by N.~Shah and
A.J.~Millis~\cite{Shah} starting from a microscopic theory involving
in-plane Green functions. An important new ingredient of the model of
Ref.~\cite{Mike} is that the local field effects are taken into account:
the changes of the phonon resonances in the infrared spectra result from
changes of the (dynamical) local electric fields associated with the
formation of the \tJPR. The same model, with a slightly different
parameterization of the electronic contribution, has been more recently
used to explain phonon anomalies in another bilayer compound, \bisccoIIz
(Bi-2212)~\cite{Zelezny}. Very recently the \tJPR\ and the related phonon
anomalies have also been observed in the trilayer compound
\biscco~\cite{Boris} and briefly discussed in terms of the Josephson
superlattice model and the local field effects. Here we present a detailed
account of several types of phonon anomalies that can occur in trilayer
cuprate superconductors (Sec.~\ref{theory}) and we compare our predictions
with experimental data for the trilayer compound \tbcco (Tl-2223) obtained
by Zetterer \etal~\cite{Zetterer} (Sec.~\ref{experiment}). Surprisingly,
these data are found to be fairly different from those of \biscco. In
Sec.~\ref{summary} we summarize our conclusions.

\section{Theory}
\label{theory}
\subsection{The basic concept and the response of a stack of
superconducting CuO$_2$ planes}
\label{basic}
A schematic representation of the three CuO$_2$ planes of a trilayer
compound and the basic structural element of Tl-2223 are shown in
Fig.~\ref{modmarel}; $E\bl$ and $E\intt$ denote the average electric
fields in the intra-trilayer and the inter-trilayer region, respectively,
and $\kappa$ ($-\kappa$) denotes the surface charge density of the lower
(upper) outer CuO$_2$ plane. The distance between the copper-oxygen planes
in the trilayer block is denoted by $d\bl$ and the distance between neighbouring
trilayers by $d\intt$. The magnitude of the inter-trilayer current density
$j\intt$ is, due to the low conductivity of this region, much smaller than
that of the intra-trilayer current density, $j\bl$. Consequently,
$j\intt$ will be neglected ($j\intt=0$). Furthermore, the current
density $j\bl$ can be expressed as $j\bl\zo=-\ii\o\eo\chi\bl\zo E\bl\zo$,
where $\chi\bl\zo$ is the local susceptibility of the Josephson junction
between the closely spaced CuO$_{\rm 2}$ layers. Because of symmetry, the
local susceptibility is the same for both regions inside the trilayer block.
As a consequence, the middle cooper-oxygen plane does not become charged
in an infrared wave; the fields and the current densities in the two
regions are the same. The trilayer block behaves as one Josephson junction
with the susceptibility $\chi\bl\zo$ connecting the outer planes of the block.

The {\it c}-axis dielectric function in the optical limit can be written
as (cf.~\cite{Ehrenreich} p. 127)
\begin{equation}
\label{eps0}
\e\zo=1+\frac{\ii}{\eo\o}\frac{j\zo}{E\zo}\;,
\end{equation}
where $j\zo$ and $E\zo$ are Fourier components of the total induced current
density along the {\it c-}axis and the total electric field, respectively.
Both variables are macroscopic, i.e., unit cell averages of the
corresponding microscopic quantities. This equation can be written as
\begin{equation}
\label{eps1}
\e\zo=\ei+\frac{\ii}{\eo\o}\frac{\sum_k\langle j_k\zo\rangle}{E\zo}\;,
\end{equation}
where $\ei$ is the interband dielectric constant, the induced current
density is expressed as the sum of microscopic
current densities $j_{\rm k}\zo$ and the volume average is explicitly
denoted by $\langle\rangle$.
For a trilayer compound, Eq.~(\ref{eps1}) can be further rewritten as
\begin{equation}
  \label{eps2}
  \e\zo=\ei+\frac{2d\bl}{2d\bl+d\intt}\frac{\ii}{\eo\o}\frac{ j\bl\zo}{ E\zo}+
  \frac{\ii}{\eo\o}\frac{\sum_k j\ph_k\zo}{E\zo}\;,
\end{equation}
where the term $2d\bl/(2d\bl+d\intt)$ is the volume fraction
of the intra-trilayer region
and $j\ph_k\zo=-\ii\o n_kQ_kr_k\zo$ represents
the (volume averaged) current density
due to vibrations of the ions of type $k$.
Further  $n_k$, $Q_k$ and $r_k\zo$ are the corresponding concentration,
effective dynamical charge, and displacement, respectively.

First we discuss the electronic part of the model, i.e., we set
$j\ph_k=0$. The electric fields $E\bl$ and $E\intt$
shown in Fig.~\ref{modmarel}\,a) are determined by the following equations
\begin{equation}
\label{Ebl}
  E\bl\zo=E'\zo+\frac{\kappa\zo}{\eo\ei}\;,
\end{equation}
\begin{equation}
\label{Eint}
E\intt\zo= E'\zo\;,
\end{equation}
\begin{equation}
\label{E}
  E\zo=\frac{2d\bl}{2d\bl+d\intt} E\bl\zo+\frac{d\intt}{2d\bl+d\intt} E\intt\zo\;,
\end{equation}
where $E'$ is the component of the average internal field $E$ that is due to
external charges and depolarization processes at high frequencies. In the
absence of the latter (i.e. for $\ei=1$) $E'$ would correspond to the field
$E_0$ of Kittel's text book~\cite{Kittel}.
The continuity equation reads
\begin{equation} \label{kappa}
  j\bl\zo=\ii\o\kappa\zo\;.
\end{equation}
Equations~(\ref{Ebl})--(\ref{kappa}) are solved by
\begin{equation}
\frac{E\intt\zo}{E\t\zo} =\frac{(2d\bl+d\intt)\,\e\bl\zo}{2d\bl\ei +
d\intt\e\bl\zo }\;,\quad
\frac{E\bl\zo}{E\t\zo}=\frac{(2d\bl+d\intt)\,\ei}{2d\bl\ei + d\intt\e\bl\zo}\;,
\end{equation}
\begin{equation}
  \e\zo=\frac{(2d\bl+d\intt)\,\ei\e\bl\zo}{2d\bl\ei +d\intt\e\bl\zo }\;,
\end{equation}
where $\e\bl\zo=\ei+\chi\bl\zo$.

In the following, the susceptibility $\chi\bl$ is taken in the form
$\chi\bl\zo=-\o\bl^2/\o^2+\ii S\bl/\o$. The first and the second term
represent the response of the superfluid and the background,
respectively\footnote{It has been recently suggested~\cite{Timusk} that the
first term should be given a finite spectral width, since it could be unrelated to the
SC condensate. The suggestion was motivated by the observation that in
strongly underdoped Y-123 the additional resonance, which is determined by
this term, appears already at temperatures much higher than $T_{\rm c}$. In
other compounds, however, it starts to appear below $T_{\rm c}$ or very
close to $T_{\rm c}$. We are convinced that at least in the superconducting
state the first term corresponds to the condensate. The origin of the
above-$T_{\rm c}$ phenomena in strongly underdoped Y-123 is not yet
clear but we believe that they are due to fluctuation effects.}.
The background is described in the simplest possible way (an infinitely
broad Drude term) in order to keep the number of parameters small.
Figure~\ref{plasmoni} shows the spectra of the real part $\sigma_1(\omega)$
of the optical conductivity $\sigma(\omega)=-\ii\o\eo\e(\o)$ for several
values of the plasma frequency $\o\bl$. Since this article aims at
explaining the experimental data for Tl-2223, the following values
characteristic for this compound have been used: $d\bl=3.2\,{\rm \AA}$,
$d\intt=11.5\,{\rm \AA}$, $A=a^2$, where $a=3.9\,{\rm \AA}$ is the in-plane
lattice parameter~\cite{Hasegawa}; further $\ei=5$ (the value typical for
the {\it c}-axis response of the high-$T_{\rm c}$ cuprates), and
$S\bl=1400\cm$ (a value, that gives a reasonable width to the \tJPR). The
maximum in the spectra of Fig.~\ref{plasmoni} corresponds to the \tJPR. It
can be seen that $\o\bl$ determines not only the frequency of the resonance
but also its spectral weight.

In order to include phonons, we have to take into account
the local fields acting on the ions participating in the vibrations.
It can be seen from Fig.~\ref{modmarel}\,a) that the charge densities
$\kappa$ and $-\kappa$ give rise to three different electric fields:
(a) the field between the trilayer blocks ($E\loc\out$),
(b) the one inside the trilayer ($E\loc\inbl$),
and (c) the field at the outer copper-oxygen planes ($E\loc\cp$).
They are given as
\begin{eqnarray}
  \label{Eout}
  E\loc\out\zo&=& E'\zo\;,\\
  \label{Ein}
  E\loc\inbl\zo&=& E'\zo+\frac{\kappa\zo}{\eo\ei}\;,\\
  \label{Ecp}
  E\loc\cp\zo&=& E'\zo+\frac{\kappa\zo}{2\eo\ei}\;.
\end{eqnarray}
As a consequence, we arrive at three different elementary types of phonons
consisting in vibrations of ions located in the three regions.

\subsection{Vibrations of ions located in the inter- or intra-trilayer region}
\label{secsimple}
In order to include a phonon mode consisting in a vibration of a given type
of ion, we have to express its contribution to the current density, $j\ph$,
and describe its influence on the fields $E\bl$ and $E\intt$. For
concreteness we focus on the inter-trilayer case; the intra-trilayer one
can be treated in an analogical way. The classical equation of motion for a
particle of mass $m$ and charge $Q$ yields the following formula
for the Fourier component of the displacement
\begin{equation}
\label{r}
r\zo=\frac{\eo\chi\zo\, E\loc\intt\zo}{nQ}\;,
\end{equation}
where $\chi$ is the polarizability of the phonon,
\begin{equation}
\chi\zo=\frac{S_Q\,\o_Q^2}{\o^2_{Q}-\o^2-\ii\o\gamma_Q}\ \hbox{with}
\quad S_Q= \frac{nQ^2}{\eo m \o_Q^2}\;.
\end{equation}
Further $n$ is the density of the vibrating ions, $n=N/[A(2d\bl+d\intt)]$,
where $N$ is the number of the ions per volume $A(2d\bl+d\intt)$; $\o_Q$ and
$\gamma_Q$ are the characteristic frequency and the broadening parameter,
respectively. The (volume averaged) current density corresponding to this
vibration is
\begin{equation}
\label{jph}
j\ph\zo=-\ii\o\eo\chi\zo\, E\loc\out\zo\;.
\end{equation}

The influence of the phonon on the electric fields can be taken into
account using the following simple model: the planes of the ions are
approximated by uniformly charged planes perpendicular to the {\it c}-axis
with the same surface charge density, $\eta=NQ/A$ (see Fig.~\ref{single}).
Since the electric field generated by a uniformly charged plane does not
depend on the distance from the plane, this assumption simplifies the model
considerably. Using Fig.~\ref{single}, we obtain the following equation for
the average field in the inter-trilayer region:
\begin{equation}
\label{Eint02}
d\intt E\intt(t)=
[d_\eta+r(t)](-E_{\eta})+
[d\intt-d_\eta-r(t)]E_{\eta}+
d\intt E'(t)\;,
\end{equation}
where $d_\eta$ is the equilibrium distance of the charged plane from the
lower CuO$_2$ plane, and $E_{\eta}=\eta/2\eo\ei$.
Screening by high frequency processes is taken into account by $\ei$ in
the denominator. Fourier transformation of Eq.~(\ref{Eint02}) yields
\begin{eqnarray}
\label{Eint2}
\nonumber
E\intt\zo&=&E'\zo-\frac{r\zo}{d\intt}\,\frac{\eta}{\eo\ei}=\\
&=&E'\zo-\frac{\gamma}{\ei}\,\chi\zo \,E\loc\out\zo\;,
\quad\gamma=\frac{2d\bl+d\intt}{d\intt}\,.
\end{eqnarray}
For the particular type of phonon Eq.~(\ref{Eint2})
substitutes for Eq.~(\ref{Eint}),
whereas Eq.~(\ref{Ebl}) remains unchanged. The second term on the right-hand
side of Eq.~(\ref{Eint2}) represents the depolarization field of the
phonon. Note that it contributes only to the field $E\intt$ and not to the
field $E\bl$. On the other hand, the volume averaged depolarization field
\begin{equation}
\frac{d\intt}{2d\bl+d\intt}\frac{\gamma}{\ei}\chi\zo E\loc\intt\zo
\end{equation}
is equal to $-P/\eo\ei$, where $P$ is the macroscopic polarization
[$P=\eo\chi\zo E\loc\intt$]. This is the textbook formula for the
macroscopic depolarization field ($E_1$ of Kittel's book) of a thin plate.
The local fields $E\loc\intt$, $E\loc\bl$, $E\loc\cp$ do not contain any
contribution of the depolarization fields. This resembles the exact
cancellation of the depolarization field and the Lorentz field ($E_2$ of
Kittel's textbook) that occurs in cubic crystals. Note that the phonon
polarizabilities $\chi$ entering the model equations represent response
functions with respect to the local fields instead of the averaged field.
In the absence of interlayer currents, the input frequencies would
correspond to the LO-frequencies while the frequencies renormalized
according to the model equations would correspond to the TO-ones;
considering only one phonon we obtain $\o_{\rm TO}=\o_Q\sqrt{1-S_Q/\ei}$.
Note that if we include the current $j\ph$ in Eq.~(\ref{eps2}), we also
have to take into account the influence of the corresponding polarization
on the electric field, i.e., the depolarization field. Otherwise the
approach would not be consistent and may lead to unphysical results.

Figures~\ref{apict}\,a) and~\ref{apict}\,b) show the real part of the
optical conductivity for two values of the phonon frequency,
$\o_{Q}=320\cm$ and $\o_{Q}=610\cm$, respectively. The values of the remaining
parameters correspond to  O2 or O3 oxygens in Tl-2223: $N=2$, $m=16$\,amu,
and we take $Q=-2|e|$ and $\gamma_{Q}=20{\rm\,cm^{-1}}$. The full lines
represent the spectra of the normal state (NS) with $\o\bl=0$ and display
only the phonon resonance. The dashed lines represent the spectra of the
superconducting state (SCS) with $\o\bl=1400\cm$ and display both the
\tJPR\ (the additional peak) and the phonon. We emphasize that the set of
the values of the parameters used to compute the SCS spectrum differs from
that of the NS only in the value of the parameter $\o\bl$. It can be
seen that if the frequency of the phonon is lower than that of the \tJPR\
[Fig~\ref{apict}\,a)], the spectral weight (SW) of the phonon is enhanced
in the SCS. If it is higher [Fig~\ref{apict}\,b)], the SW is reduced. Note
that the qualitative characteristics of the anomaly depend only on the
order of the two frequencies.

Next we consider a phonon consisting in a vibration
of intra-trilayer ions of a given type.
The derivation of the formula for the depolarization field is entirely
analogical to the case of the inter-trilayer ions. We obtain
\begin{eqnarray}
\nonumber
E\bl\zo&=&E'\zo+\frac{\kappa\zo}{\eo\ei}-
\frac{r\zo}{2d\bl}\,\frac{\eta}{\eo\ei}=\\
  \label{Ebl2}
&=&E'\zo+\frac{\kappa\zo}{\eo\ei}-
\frac{\delta}{\ei}\,\chi\zo \,E\loc\inbl\zo\;,
\quad\delta=\frac{2d\bl+d\intt}{2d\bl}\;.
\end{eqnarray}
This equation replaces Eq.~(\ref{Ebl}) whereas Eq.~(\ref{Eint})
remains unchanged. The main difference from the inter-trilayer
case is that the ions feel the local field $E\loc\inbl$ instead of
$E\loc\out$. Figures~\ref{apict}\,c) and~\ref{apict}\,d) show, how the
phonon resonance is influenced by the superconducting transition. The
values of the phonon frequencies are: $\o_Q=320\cm$ and $610\cm$,
respectively. The values of the other parameters correspond to Ca in
Tl-2223: $N=2$, $m=40$\,amu, and we take  $Q=2|e|$ and $\gamma_Q=3{\rm\,cm^{-1}}$. It
can be seen that if the phonon  frequency is lower than that of the
\tJPR\ [Fig~\ref{apict}\,c)], its SW is reduced and the phonon softens
considerably. If it is higher [Fig~\ref{apict}\,d)], the SW is enhanced and
the phonon frequency increases. Note that the phonons are renormalized
already in the NS because of the background term in the formula for
$\chi\bl$. The renormalization manifests itself, among others, in
a non-Lorentzian shape of the resonances.

We have shown in Sec.~\ref{basic}  that the trilayer block behaves simply as
a bilayer one with the interplanar distance of $2d\bl$. The approach
presented here can thus be applied to both trilayer and bilayer compounds:
in the latter case, $2d\bl$ in the coefficients $\gamma$ and $\delta$ has
to be substituted by $d\bl$ and we have $\gamma=(d\bl+d\intt)/d\intt$ and
$\delta=(d\bl+d\intt)/d\bl$.

\subsection{Vibrations of ions located in the outer CuO$_2$ planes of the
trilayer block}
\label{secplane}
For concreteness we further discuss the infrared-active (in-phase)
vibrations of the oxygens O1 of the outer CuO$_{2}$ planes. Vibrations of the
copper ions, however, can be treated in the same way. Boundaries between
the intra- and inter-trilayer regions can be identified either with the
static planes of the Cu ions or with the planes of the vibrating oxygens.
In order to simplify our considerations we choose here the first approach.
We have checked that for reasonable values of effective charges the use of
the second one leads to very similar results. When expressing the
depolarization fields in any of these approaches we encounter a problem:
the interface between the inter- and the intra-trilayer regions is
repeatedly crossed by a charged plane, which makes the time dependence of
the fields $E\bl$ and $E\intt$ unharmonic. This difficulty results from our
approximation of the ionic planes by two-dimensional homogeneously charged
planes. In order to get rid of the problem and keep the useful
approximation at the same time, we represent the planar oxygens O1 by two
homogeneously charged planes as shown in Fig.~\ref{plane}.

The surface charge density of each vibrating plane is
$\eta/2$ with $\eta=N_{\rm O1}Q_{\rm O1}/A$, $N_{\rm O1}=2$.
Following the derivation of the preceding section,
we obtain the subsequent formula for the average field
between the upper two CuO$_2$ planes of the trilayer:
\begin{eqnarray}
\nonumber
E_{\rm bl}\zo&=&E'\zo+\frac{\kappa\zo}{\eo\ei}-
  \frac{r\zo}{d\bl}\frac{\eta/2}{\eo\ei}=\\
\label{Ebl3}
&=&E'\zo+\frac{\kappa\zo}{\eo\ei}-
  \frac{2\alpha}{\ei}\,\chi_{\rm O1}\zo\, E\loc\cp\zo\;,
\quad\alpha=\frac{2d\bl+d\intt}{4d\bl}\,.
\end{eqnarray}
We have used the relation between the displacement $r$
and the local field $E\loc\cp$,
$r=\eo\chi_{\rm O1} E\loc\cp/(n_{\rm O1}Q_{\rm O1})$ [cf. Eq.~(\ref{r})],
where $n_{\rm O1}$
is the concentration of the oxygens O1 of an upper CuO$_2$ plane,
$n_{\rm O1}=N_{\rm O1}/[A(2d\bl+d\intt)]$. The average field between the
two lower CuO$_{2}$ planes is also given by Eq.~(\ref{Ebl3}). Consequently,
the central CuO$_2$ plane remains without time dependent charge density
even for this type of vibration. The equation for the field $E\intt$,
\begin{eqnarray}
  \label{Eint3}
\nonumber
E\intt\zo&=&E'\zo- \frac{r\zo}{d\intt}\frac{\eta}{\eo\ei}=\\
&=&E'\zo-\frac{2\beta}{\ei}\,\chi_{\rm O1}\zo E\loc\cp\zo\;,
\quad\beta=\frac{2d\bl+d\intt}{2d\intt}
\end{eqnarray}
can be obtained in an analogical way. Equations~(\ref{Ebl3}) and~(\ref{Eint3})
substitute for~(\ref{Ebl}) and~(\ref{Eint}).
The current density due to the vibrating oxygens is given as the sum of the contributions
of the two outer oxygen planes [cf. Eq.~(\ref{jph})]
\begin{equation}
\label{jph3}
j\ph(\omega)= -2\ii\o\eo\chi_{\rm O1}(\omega)E\loc\cp(\omega)\;.
\end{equation}
The formulas for $\alpha$ and $\beta$ of a bilayer system can be obtained
along the lines of the considerations at the end of the previous
subsection: $\alpha=(d\bl+d\intt)/2d\bl$ and
$\beta=(d\bl+d\intt)/2d\intt$.

Note that the second approach (with the boundaries associated with the
charged planes of the planar oxygens) yields slightly different expressions
for $\alpha$ and $\beta$ in Eqs.~(\ref{Ebl3}) and (\ref{Eint3}) and also
the formula for the current density due to the phonon differs from
Eq.~(\ref{jph3}), see Appendix. In the previous works concerning bilayer systems
Y-123~\cite{Mike} and Bi-2212~\cite{Zelezny,Mike2}, $\alpha$ and $\beta$
have been calculated using the second approach but $j\ph$ using the first
one, i.e., Eq.~(\ref{jph3}). Fortunately, the errors caused by this formal
inconsistency are negligible because the values of $\alpha$ and $\beta$
resulting from both approaches are very close.
Note that there are some errors in the third line of Table~1 of
Ref.~\cite{Mike}, the correct values will be published in an
erratum~\cite{erratum}.

Figure~\ref{ocp}\,a) and~\ref{ocp}\,b) shows the interplay between the
\tJPR\ and the phonon for two values of the phonon frequency: $\o_{\rm
O1}=400\cm$ and $660\cm$. The values of the remaining parameters are $N_{\rm
O1}=2$, $m_{\rm O1}=16$\,amu, and we take $Q_{\rm O1}=-2|e|$ and
$\gamma_{\rm O1}=20{\rm\,cm^{-1}}$. If the phonon frequency is lower
than that of the \tJPR\ [Fig~\ref{ocp}\,a)], the formation of the \tJPR\ is
associated with a reduction of the phonon SW and its softening by about
$5\cm$. An even more pronounced phonon anomaly of this type occurs for the
bond bending mode in underdoped Y-123~\cite{Mike}. The fact, that the
anomaly expected to occur in Tl-2223 is smaller than in Y-123, is connected
to the difference in the width of the multilayer block. If the phonon
frequency is higher than that of the \tJPR\
[Fig~\ref{ocp}\,b)], the SW of the phonon is enhanced in the SCS.

\subsection{Composite vibrations}
\label{seccomposed}
Up to this point we have described elementary vibrations where only one
type of ion participates. This can be a good approximation for a mode where
the major part of polarization is due to vibrations of one type of ion. For
modes where several types of ions contribute significantly an extended
version of the model is required. In this section we study composite modes
of the oxygens of copper-oxygen planes. According to the results of the shell
model calculations~\cite{Kulkarni}, in trilayer compounds such modes can be
expected to occur.

In order to treat the composite modes as simply as possible we introduce
the mechanical coupling scheme shown in Fig.~\ref{pruzina}, which provides an
out-of-phase and an in-phase resonance. The equations of motion read
\begin{eqnarray}
m_{\rm O}\frac{\d^2 r(t)}{\d t^2}&=&
-k_1 r(t)+k_2[z(t)-r(t)]-
m_{\rm O}\gamma_{\rm O1}\frac{\d r(t)}{\d t}+F_r(t)\;,\\
m_{\rm O}\frac{\d^2 z(t)}{\d t^2}&=&\
 2k_2[r(t)-z(t)]-
m_{\rm O}\gamma_{\rm O4}\frac{\d z(t)}{\d t}+F_z(t)\;.
\end{eqnarray}
The forces $F_r$ and $F_z$ are proportional to the local fields:
$F_r=E\loc\cp Q_{\rm O1}$ and $F_z=E\loc\inbl Q_{\rm O4}$.
For the Fourier components of the ionic displacements we obtain
\begin{equation}
  r\zo=\frac{F_r\zo\, q_z\zo+F_z\zo\, k_2}{q_r\zo\, q_z\zo-2k_2^{2}}\;,
  \quad
  z\zo=\frac{F_z\zo\, q_r\zo+2F_r\zo\, k_2}{q_r\zo\, q_z\zo-2k_2^{2}}\;.
\end{equation}
Here $q_r=-\o^2m_{\rm O}+k_1+k_2-\ii\o\gamma_{\rm O1}m_{\rm O}$,
$q_z=-\o^2m_{\rm O}+2k_2-\ii\o\gamma_{\rm O4}m_{\rm O}$.
The depolarization field for a composite vibration
is the sum of the contributions of the elementary vibrations.
Following the consideration of the previous subsections we obtain the
equations for the mean fields:
\begin{eqnarray}
  \label{Ebl4}
  E\bl\zo&=& E'\zo+\frac{\kappa\zo}{\eo\ei}-
  \frac{r\zo\,\eta_{\rm O1}+z\zo\,\eta_{\rm O4}}{2d\bl\eo\ei }\;,\\
  \label{Eint4}
  E\intt\zo&=& E'-\frac{r\zo}{d\intt}\frac{\eta_{\rm O1}}{\eo\ei}\;,
\end{eqnarray}
where $\eta_{\rm O1}=N_{\rm O1}Q_{\rm O1}/A$
and $\eta_{\rm O4}=N_{\rm O4}Q_{\rm O4}/A$.
Equations~(\ref{Ebl4}) and~(\ref{Eint4})
again replace Eqs.~(\ref{Ebl}) and~(\ref{Eint}). The current
density due to the vibrations of the oxygens is
\begin{equation}
\label{proudslozeny}
\sum_k j\ph_k\zo=-2\ii\o n_{\rm O1}Q_{\rm O1}r\zo-\ii\o n_{\rm O4}Q_{\rm
O4}z\zo\;.
\end{equation}
The first and the second term on the right side of the
Eq.~(\ref{proudslozeny}) is due to vibrations of oxygens in the outer and
inner copper-oxygen planes, respectively.

Figure~\ref{twin} shows the real part of the optical conductivity
for the model described above.
The values of the parameters are:
$N_{\rm O1}=N_{\rm O4}=2$, $Q_{\rm O1}=Q_{\rm O4}=-2|e|$, $m_{\rm O}=16$,
and we have chosen $\gamma_{\rm O1}=\gamma_{\rm O4}=20{\rm\,cm^{-1}}$,
$k_{2}/k_{1}=0.3$ and $\omega_{1}=\sqrt{k_{1}\over m}=320{\rm\,cm^{-1}}$
(a), $600{\rm\,cm^{-1}}$ (b), $800{\rm\,cm^{-1}}$ (c), and
$1100{\rm\,cm^{-1}}$ (d).
The model yields an out-of phase (OPR) resonance and an
in-phase (IPR) one. If the OPR frequency is lower than that of the
\tJPR\ [Fig.~\ref{twin}\,a)], its SW increases considerably when going from
the normal to the SCS. On the contrary, if its frequency is higher
[Fig.~\ref{twin}\,b)], the SW of the phonon decreases. In both cases the
phonon frequency remains approximately constant under the superconducting
transition. For the IPR the situation is different: if its frequency is
lower than that of the \tJPR\ [Fig.~\ref{twin}\,b) and~c)], its SW is
strongly reduced and its frequency decreases. If the IPR frequency
is higher [Fig.~\ref{twin}\,d)], the changes are reverse: the SW increases
and the phonon hardens. In Fig.~\ref{twin}\,a) the interplay of the
depolarization fields and the electronic background leads to a strong
overdamping of the IPR. In Figs.~\ref{twin}\,c) and d), the OPR is located
above the displayed frequency range. It can be seen that the IPR frequency
is renormalized much more than that of the OPR. This can be
understood: for the IPR the displacements of the oxygens $r$ and $z$ have
the same sign, therefore the depolarization field (the third term on the
right-hand side of Eq.~(\ref{Ebl4}), determining the frequency shift,
is large. On the contrary, for the OPR the signs of the displacements
are opposite and the depolarization field is therefore small. Note that the
qualitative aspects of the phonon anomalies of the composite modes are
insensitive to changes of the value of the ratio $k_1/k_2$.

The changes of the spectral weights reported in this section can be
elucidated in terms of the changes of the local fields. Figure~\ref{local2}
displays frequency dependencies of the ratios $E\loc\inbl/E$, $E\loc\out/E$
and $E\loc\cp/E$ in the SCS obtained using Eqs.~(\ref{Ebl})--(\ref{Ecp}).
The phonons have not been taken into account and $S_{bl}=0$. The
singularity at $\o_{\rm p}\approx 500\cm$ corresponds to the \tJPR. Since
the values of the ratios in the NS  are equal to one, their values in the
SCS express the relative changes of the local fields when going from the NS
to the SCS. Let us consider, e.g., the local field $E\loc\out$ that
acts on ions situated in the inter-trilayer region. If the frequency $\o_Q$
of a phonon involving vibrations of these ions is lower than $\o_{\rm p}$,
the local field $E\loc\out(\o_Q)$ increases and the SW of
the phonon increases [cf. Fig~\ref{apict}\,a)]. If the frequency is higher,
$E\loc\out(\o_Q)$ decreases and so does the spectral weight
[cf.~Fig~\ref{apict}\,b)]. The SW changes of phonons consisting in a
vibration of intra-trilayer ions or ions in the outer CuO$_2$ planes can be
elucidated in a similar way. In order to understand the SW changes for a
composite vibration of the planar oxygens, we have to take into account
changes of the absolute values of the fields, changes of their relative
signs, and the displacement pattern. In the frequency region $\o_Q<\o_{\rm
p}$, the local field $E\loc\inbl$ changes sign under the superconducting
transition while the sign of $E\loc\cp$ remains unchanged, i.e., in the
SCS the local fields are antiparallel in contrast to the NS, where they are
parallel. Considering the out-of-phase resonance, the eigenvector pattern
is in disagreement with that of the local field in the NS while they are in
accord in the SCS. Consequently, the amplitude of the vibration is enhanced
under the superconducting transition as well as the SW of the phonon [cf.
Fig.~\ref{twin}\,a)]. For the IPR the situation is reversed,
i.e., the two patterns are in agreement in the NS while they disagree in
the SCS. As a consequence, the SW of the phonon resonance decreases [cf.
Fig.~\ref{twin}\,c)]. For $\o_{\rm Q}>\o_{\rm p}$, the two local fields are
parallel in both NS and SCS, and upon entering the SCS, their
amplitudes increase. Since the eigenvector pattern of the in-phase
resonance agrees with that of the electric field, its SW increases [cf.
Fig.~\ref{twin}\,d)].

The shifts of the phonon frequencies are related to the depolarization
fields. Since the local fields shown in Fig.~\ref{local2} have been
obtained within a model, that does not incorporate the phonons, i.e., does
not contain the depolarization fields, this picture cannot explain the shifts.

\section{Comparison with experimental data and discussion}
\label{experiment}
Pronounced phonon anomalies have been observed
in the far-infrared spectra of the trilayer compound Tl-2223~\cite{Zetterer}.
In this section we discuss the experimental data,
present an assignment of the phonon modes,
and compare the experimental results with our predictions.
Finally we comment on the phonon anomalies
in Bi$_{2}$Sr$_{2}$Ca$_{2}$Cu$_{3}$O$_{10}$ (Bi-2223).

\subsection{\tbcco}
\label{tbcco}
The conductivity spectra of Tl-2223 shown in Fig.~\ref{vodiv} contain a
structure situated on a high conductivity background. The spectra were
obtained~\cite{Zetterer} by Kramers-Kronig transformation of reflectivity
data of a ceramic sample consisting of microcrystals with
random orientations. It is known, that sharper structures in the data of
polycrystalline samples of high-$T_{\rm c}$ compounds corresponds largely to
the $c$-axis polarized phonons and that the metallic background originates
predominantly from the {\it a-b} conductivity (see \cite{Litvinchuk} and
Ref. therein). The absence of {\it a-b} polarized phonons can be seen,
e.g., by comparing the reflectivity spectra of a ceramic Y-123 ceramic
sample~\cite{Genzel} with those of a $c$-axis oriented
monocrystal~\cite{Homes}. The phonon structure (see Fig.~\ref{vodiv})
exhibits several changes under the superconducting transition: the SW of
the phonon at $580\cm$ decreases, a wide additional absorption band appears
in the frequency region around $500\cm$, the SW of the phonon at $370\cm$
increases, the frequency of the phonon at $305\cm$ decreases by about
$20\cm$, the SWs of the four low frequency phonons increase.

The lattice dynamical study of Tl-2223~\cite{Kulkarni} predicts eight
infrared-active $c$-axes polarized phonons with eigenvectors shown in
Fig.~\ref{thtisk}. Only seven phonon resonances, however, can be found in
the normal state spectra of Fig.~\ref{vodiv}\,a). We suggest that the
phonon, that is not seen in the data, is the one with the predicted TO
frequency of $443\cm$ because we do not find any phonon around $443\cm$ in
the normal state spectra and because this mode should have a
rather low oscillator strength. Having identified the missing phonon, we
base the assignment of the phonons on the order of the TO frequencies. The
differences between the measured and the predicted TO frequencies are
within 20\% which can be considered as a good agreement. Note however that
also the results of the shell model calculations should be taken with some
precaution. For example, the calculations do not consider free charge
carriers.

As discussed in Sec.~\ref{theory}, the temperature evolution of a phonon
resonance depends on whether its frequency is higher or lower than that of
the \tJPR. We shall therefore identify the \tJPR\ in the conductivity
spectrum and subsequently explain the observed changes of the phonons. We
suggest that the additional absorption band appearing in the SCS around
$500\cm$ corresponds to the \tJPR. The band is rather wide (more than
$100\cm$), similarly as in underdoped Y-123, and, in addition, located in
the frequency region of no phonon related feature in
the normal state data.

The major part of the polarization of the phonon at $580\cm$ (the $583\cm$
mode in Fig.~\ref{thtisk}) is due to the oxygens in the Tl-O layers.
Vibrations of inter-trilayer ions have been described in
Sec.~\ref{secsimple}, the corresponding model spectrum with the appropriate
value of the phonon frequency is shown in Fig.~\ref{apict}\,b). The
calculation reproduces the experimentally observed decrease of the SW.
Note, however, that the absolute values of the calculated and the measured
conductivities cannot be compared because of the polycrystalline structure
of the sample. In addition, in the model spectra of the SCS the phonon line
appears as a shoulder on the background of the \tJPR; furthermore a week
broad band emerges at even higher frequencies. Both features seem to be
present also in the experimental data. The latter can be interpreted as a
part of the \tJPR, which is separated from the main part by the phonon.
This feature is even more pronounced in the reflectivity data of the
bilayer system Tl$_{2}$Ba$_{2}$CaCu$_{2}$O$_{8}$~\cite{Renk,Renk2}. In the
low temperature spectrum of this compound the \tJPR\ is centered at about
$550\cm$ but a part of it appears also around $700\cm$, above the phonon at
$600\cm$.

The major part of the polarization of the $370\cm$ mode of Fig.~\ref{vodiv}
is due to an out-of-phase vibration of the oxygens of the outer and the
inner copper-oxygen planes (see the $383\cm$ mode in Fig.~\ref{thtisk}). In
section~\ref{secplane} we have described vibrations of the oxygens in the
outer CuO$_{2}$ planes, the vibrations of those of the inner plane can be
studied along the lines of Sec.~\ref{secsimple}. The corresponding model
spectra are shown in Fig.~\ref{ocp}\,a) and Fig.~\ref{apict}\,c),
respectively. Both calculations yield a decrease of the SW and a softening
of the phonon, in contrast with the experimental observation. However, for
composite (out-of-phase) vibrations our model yields a SW increase without
any frequency change [see Fig.~\ref{twin}\,a)], in agreement with the
experimental data. This agreement supports the local field picture
discussed in the end of Sec.~\ref{seccomposed}. In particular, it
demonstrates dramatic spatial variations of the local electric field, even
its sign changes on the scale of the inter-plane distance.

The polarization pattern corresponding to the phonon at $305\cm$ ($292\cm$
mode in Fig.~\ref{thtisk}) is more complicated. The major part of its
polarization is due to vibrations of the Ca ions and the apical oxygens. In
section~\ref{secsimple}, we have described vibrations of intra-trilayer and
inter-trilayer ions, the corresponding model spectra are shown in
Fig.~\ref{apict}\,c) and Fig.~\ref{apict}\,a), respectively. Both exhibit
a softening of the phonon, the former one shows in addition a SW
decrease, similar to the one occurring in the experimental data. This
suggests that the contribution of the Ca ions to the mode is more
pronounced. The model spectrum of Fig.~\ref{apict}\,c) shows a softening of
about $40\cm$, much larger than the experimental value of $20\cm$.
It is possible that the presence of other phonons in a more complete model
would yield a lower value of the frequency shift. The SW predicted by the
model is smaller than the experimental one, both in the NS and SCS. A
better agreement could be probably obtained by considering a more complex
polarization diagram. We have also to keep in mind the possibility that the
mode at $305\cm$ corresponds to the in-phase mode of the planar oxygens
(the $443\cm$ mode in Fig.~\ref{thtisk}), i.e., that the order of the
in-phase and the out-of-phase mode is different from what is suggested by
the shell model calculations. This seems to be the case in
Bi-2223~\cite{Boris}. The corresponding model spectrum shown in
Fig.~\ref{twin}\,b) exhibits a remarkable softening of the phonon resonance
and the SW decrease as well.

Since the ceramic sample consist of randomly oriented microcrystals of the
highly anisotropic material, the interpretation of the data is not
straightforward. It is important to find out whether the above discussed
spectral changes are intrinsic features of the crystal or artefacts
induced by the polycrystalline structure. The reflectivity of
polycrystalline Tl$_2$Ba$_2$CuO$_6$ with one CuO$_2$ layer per formula unit
(see Fig.~12 in Ref.~\cite{Renk}) exhibits only very small differences
between above and below $T_{\rm c}$, but the structures over $250\cm$ in the
data of the double layer compound (Tl$_{2}$Ba$_{2}$CaCu$_{2}$O$_{8}$)
and the triple layer compound (Tl-2223) change considerably (see Fig.~13
and 15 of Ref.~\cite{Renk}). We are therefore convinced that the main
changes of the phonon structures are intrinsic effects rather than
artifacts. However, some spectral characteristics must be influenced by the
polycrystalline structure. We suggest that it is the reason of the unexpectedly
large values of the SW of phonons~\cite{Zetterer} and the
anomalies of the three low frequency modes.

\subsection{A remark on the phonon anomalies in \biscco}
Recently A.~V.~Boris \etal~\cite{Boris} reported the far-infrared {\it
c}-axis spectra of the trilayer compound Bi-2223, that has the same crystal
structure as Tl-2223. It is interesting to find out whether the phonon
anomalies of this compound can also be explained by using the model
introduced above, especially those of the two modes that are specific to
the trilayer compounds: the in-phase and the out-of phase oxygen bond
bending modes. A qualitative explanation of the SW changes has
already been presented in Ref.~\cite{Boris}. It is further interesting to
explore the differences between the spectra of the two compounds.

In the SCS spectra of Bi-2223, a broad additional absorption band appears
around $500\cm$ (see Fig.~1 of Ref.~\cite{Boris}) which has been
attributed to the \tJPR. Results of the shell model calculations suggest
that the spectral structures at $360\cm$ and at $400\cm$ correspond to the
in-phase and the out-of-phase oxygen bond-bending modes, respectively. The
SW of the former decreases in the SCS, in contrast to the SW of the latter,
that increases below $T_{\rm c}$. The two features are in qualitative
agreement with our predictions [see Figs.~\ref{twin}\,c) and \ref{twin} a),
respectively]. The observed softening of the phonon at $360\cm$ (about $5\cm$,
the value of the shift is doping dependent), however, is considerably
smaller than in our model calculations. It may be due to a coupling between
the phonon and the pronounced mode at $305\cm$, mediated by the
depolarization fields, that is not included in our model.

The main differences between the spectra of Tl-2223 and  Bi-2223
are: (a) The spectra of Tl-2223 exhibit only three phonon related
structures at frequencies above $250\cm$. This is in contrast with Bi-2223,
where four main phonon bands have been found. (b) The mode at $305\cm$ in
Bi-2223 has a large SW and its contribution to the spectra is almost
temperature independent. No such mode exists in Tl-2223. (c) The spectral
weight of the \tJPR\ in Tl-2223 seems to be much larger than in Bi-2223 as
described below.

Measurements on a single-crystal allowed the authors of Ref.~\cite{Boris}
to obtain the SW of the \tJPR\ as the difference between the integrated SCS
and NS spectra. For a Bi-2223 sample with $T_{\rm c}=102\K$ that has
approximately the same frequency of the \tJPR\ as the Tl-2223 sample
studied in \cite{Zetterer}, the SW of the \tJPR\ is approximately
$1100\,\Omega^{-1}\,{\rm cm}^{-2}$. Concerning Tl-2223, we estimate that
the SW is about $8000\ \Omega^{-1}\,{\rm cm}^{-2}$ from the area between
the 115\K\ and the 25\K\ spectra as shown in Fig.~\ref{vodiv}\,d). The SW
in Tl-2223 is thus several times larger than in Bi-2223. It must be of
course affected by the ceramic nature of the sample, nevertheless, it can
be seen even by comparison with the strongest phonons in the spectrum that
the SW is fairly large. The SW of the \tJPR\ predicted by our model is
$18\,000\ \Omega^{-1}\,{\rm cm}^{-2}$. This prediction is about twice as
much as the observed value for Tl-2223, but the value for Bi-2223 is an
order of magnitude lower. We emphasize, that the SW of the \tJPR\ in our
model is not an arbitrary parameter but is determined by the frequency of
the \tJPR\ (see Fig.~\ref{plasmoni}). The reason why the infrared spectra
of Tl-2223 and of Bi-2223 differ so much is not yet understood and
remains as an open question for future investigations.

\section{Summary and conclusions}
\label{summary}
We have extended the model of Ref.~\cite{Mike} that accounts for
phonon anomalies occurring in bilayer compounds to trilayer high-$T_{c}$
superconductors and discussed several possible phonon anomalies that can
occur in these systems. The model  explains the experimental
infrared data of the trilayer high-$T_{\rm c}$ compound Tl-2223 obtained by
Zetterer {\it et al.}~\cite{Zetterer}. The broad maximum that appears in the conductivity
spectra below $T_{\rm c}$ and represents the dominant feature of the low
temperature spectra corresponds to the transverse Josephson plasma
resonance. The phonon anomalies can be explained in
terms of changes of the local electric fields induced by the formation of
the resonance. Of particular interest is the anomaly of the $370\cm$ mode,
i.e., the large increase of its spectral weight below $T_{\rm c}$. We recall
that this mode consist in an out-of-phase vibration of the planar oxygens and
is thus specific to the trilayer compounds. Its anomalous behavior is due
to the fact that the displacement pattern does not agree with the pattern
of the local field in the normal state, but it does agree with the one of
the superconducting state, where the sign of the field changes from the
inner CuO$_{2}$ plane to the outer one.

The qualitative and in some cases even quantitative agreement between the
predictions of our model and the experimental data indicates that the model
captures well the underlying mechanism. According to the model, the
anomalies are caused by a change of the electronic ground state associated
with a crossover of the {\it c}-axis dynamics from incoherent to coherent,
the change which would presumably occur even in the absence of any
electron-phonon coupling. It seems thus that the anomalous infrared-active
phonons are passive rather than active players in the mechanism of
superconductivity.

The spectra of Tl-2223 are qualitatively similar to those of Bi-2223
reported by Boris {\it et al.}: both exhibit the additional absorption band
and the anomalies of three high-frequency phonons. The magnitudes of the
effects, as far as we can conclude based on the data of polycrystalline
Tl-2223, are considerably different. This suggests that the
spacing layers of the two materials possess fairly different electronic
properties.
\section*{Acknowledgements}
We acknowledge discussions with J.~Huml\'{\i}\v{c}ek and C.~Bernhard.
The work has been supported by the project MSM 143100002 of the Ministry of
Education of Czech Republic.

{
\appendix
\section{Appendix}
Here we present the formulas of the approach where boundaries are
associated with the charged planes of the planar oxygens.
Equations~(\ref{Ebl3}) and~(\ref{Eint3}) remain as they are with
\begin{equation}
  \alpha=-\left(\frac{\eta_{\rm tl}+\eta_{\rm Cu}}{\eta_{\rm O}}\right)
  \frac{2d\bl+d\intt}{4d\bl}\,
\end{equation}
\begin{equation}
  \beta=-\left(\frac{\eta_{\rm int}+\eta_{\rm Cu}}{\eta_{\rm O}}\right)
  \frac{2d\bl+d\intt}{4d\intt}\,
\end{equation}
where $\eta_{\rm tl}$ ($\eta_{\rm int}$) is the total surface charge
density of the intra-trilayer (inter-trilayer) region, i.e., the sum of the
densities of the charged planes located inside this region. Further
$\eta_{\rm Cu}$ ($\eta_{\rm O}$) is the surface charge density of the Cu
ions (O ions)  of a CuO$_2$ plane. Equation~(\ref{jph3}) is substituted by
\begin{equation}
  j\ph\zo=-2\ii\o\eo\chi_{\rm O1}\zo E^{\rm loc*}\cp\zo
\end{equation}
with
\begin{equation}
E^{\rm loc*}\cp\zo=
E'+\frac{\kappa\zo}{2\eo\ei}\frac{4d\bl}{2d\bl+d\intt}\,\alpha\;.
\end{equation}
The formulas for $\alpha$, $\beta$, and $E^{\rm loc*}\cp$ of a bilayer
system can be obtained by substituting $2d\bl$ by $d\bl$.
}



\newpage
Figure captions:
\begin{figure}[h]
\caption{a) Schematic representation of the model of a trilayer superconductor.
The CuO$_{\rm 2}$ planes are approximated by homogeneously charged planes.
The symbol $\kappa$ ($-\kappa$) denotes the surface charge density of the upper
(lower) CuO$_2$ plane.
b) The basic structural element of \tbcco~\cite{Hasegawa}.
The oxygens in the outer (inner) CuO$_{2}$ planes are labelled as O1 (O4),
the oxygens in the BaO (TlO) planes as O2 (O3).}
\label{modmarel}
\end{figure}
\begin{figure}[h]
\caption{Real part of the {\it c}-axis conductivity
of the electronic part of the model for the following values
of the intra-trilayer plasma frequency $\o\bl$:
$0$ (normal state), $400,\ 700,\ 1000$ and $1400\cm$.
It can be seen that $\o\bl$ determines
not only the frequency of the transverse Josephson plasma resonance,
but also its spectral weight.}
\label{plasmoni}
\end{figure}

\begin{figure}[h]
\caption{Schematic representation of our approach to treat
the influence of the phonon polarization on the electric field.
The plane of ions of a given type
is approximated by the uniformly charged plane
with the same charge per unit area $\eta$.
The displacement of the plane from its equilibrium position (dashed line)
is denoted by $r$.
The electric field due to the charged plane
is labelled as $\pm E_{\rm \eta}$, $E_{\eta}=\eta/(2\eo\ei)$.}
\label{single}
\end{figure}

\begin{figure}[h]
\caption{Anomalies of a phonon consisting
in a vibration of ions situated in the inter-trilayer region
[a) and b)] or in
the intra-trilayer region [c) and d)].
The solid (dashed) lines correspond to the normal (superconducting) state.
The transverse Josephson plasma resonance is labelled as t-JPR
and the phonon resonance
as P$_{\rm O3}$ and P$_{\rm Ca}$
in Figs. a), b) and c), d), respectively.
The values of the parameters used in the computations
are given in the text.}
\label{apict}
\end{figure}

\begin{figure}[h]
\caption{The scheme used for the evaluation of the depolarization fields of the
oxygen-bond bending mode. The oxygens of an outer CuO$_{2}$ plane are
represented by two homogeneously charged planes.}
\label{plane}
\end{figure}

\begin{figure}[h]
\caption{Anomalies of a phonon consisting in vibrations of the oxygens
of the outer CuO$_{2}$ planes.
The dashed (solid) lines correspond to the normal (superconducting) state.
The transverse Josephson plasma resonance is labelled as t-JPR
and the phonon resonance as P$_{\rm O1}$.
The values of the parameters used in the computations
are given in the text.}
\label{ocp}
\end{figure}

\begin{figure}[h]
\caption{String model used to describe the composite vibrations
of the planar oxygens.
The displacement of the oxygens O1 (O4) from their equilibrium position
is labelled as $r$ ($z$),
the string constants are denoted by $k_1$ and $k_2$.
The local electric fields $E\loc\cp$ ($E\loc\bl$) act on the oxygens O1
(O4).
}
\label{pruzina}
\end{figure}

\begin{figure}[h]
\caption{
Anomalies of the in-phase resonance (IPR)
and the out-of-phase resonance (OPR)
of the planar oxygens.
The dashed (solid) lines correspond to the normal (superconducting) state.
The transverse Josephson plasma resonance is labelled as t-JPR.
The values of the parameters used in the computations
are given in the text.}
\label{twin}
\end{figure}

\begin{figure}[h]
\caption{
Frequency dependencies of the normalized local electric fields in the
superconducting state.
The symbols  $E\loc\intt$, $E\loc\inbl$ and $E\loc\cp$
denote the local fields in the inter-trilayer region,
inside the trilayer,
and on the outer cuprate planes, respectively.
The normalized local fields
acting on the ions participating in the modes
at 305, 380 and $580\cm$,
are indicated by A, B, and C, respectively. }
\label{local2}
\end{figure}

\begin{figure}[h]
\caption{Conductivity spectra of \tbcco\ ($T_{\rm c}=112\K$) adapted from
Ref.~\cite{Zetterer}. The spectra have been obtained by Kramers-Kronig
transformation of reflectivity data of a ceramic sample. Part d) contains
the spectra of part b) and c). The shaded area between the 115\K\ and the
25\K\ spectra gives us an rough estimate of the spectral weight of the transverse
Josephson plasma resonance.}
\label{vodiv}
\end{figure}

\begin{figure}[h]
\caption{Eigenvector diagrams of the {\it c}-axis polarized infrared-active
phonons of \tbcco\ with the corresponding TO (LO) frequencies
obtained by shell model calculations in Ref.~\cite{Kulkarni}.}
\label{thtisk}
\end{figure}

\clearpage
Dubroka \etal\
Fig.~\ref{modmarel}\\[2cm]
  \begin{center}
  \includegraphics[width=16cm]{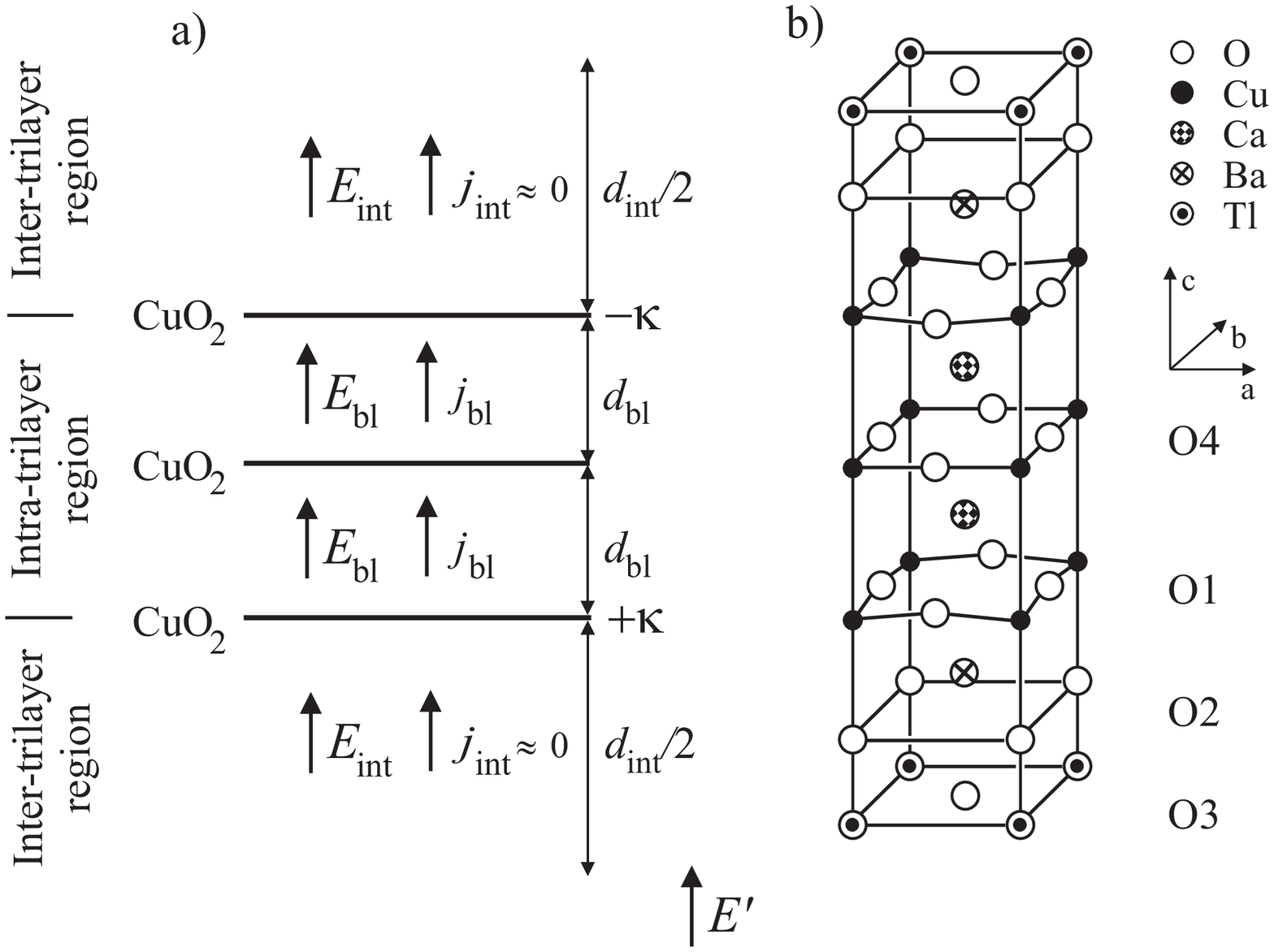}
  \end{center}

\newpage
Dubroka \etal\
Fig.~\ref{plasmoni}  \\[2cm]
  \begin{center}
  \includegraphics[width=10cm]{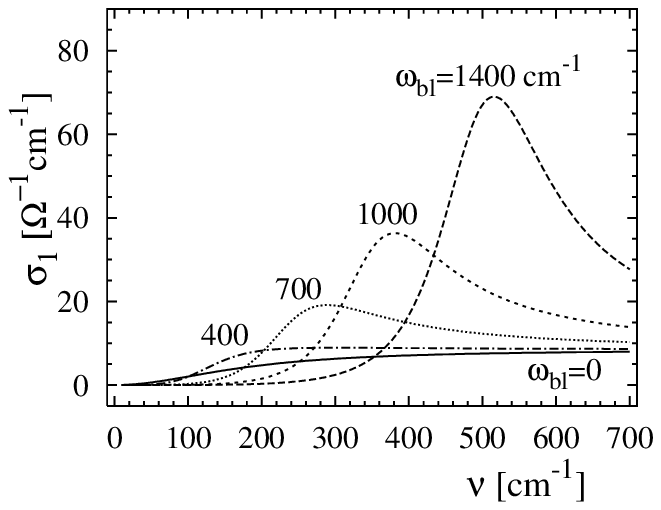}
  \end{center}

\newpage
Dubroka \etal\
Fig.~\ref{single}\\[2cm]
  \begin{center}
  \includegraphics[width=8cm]{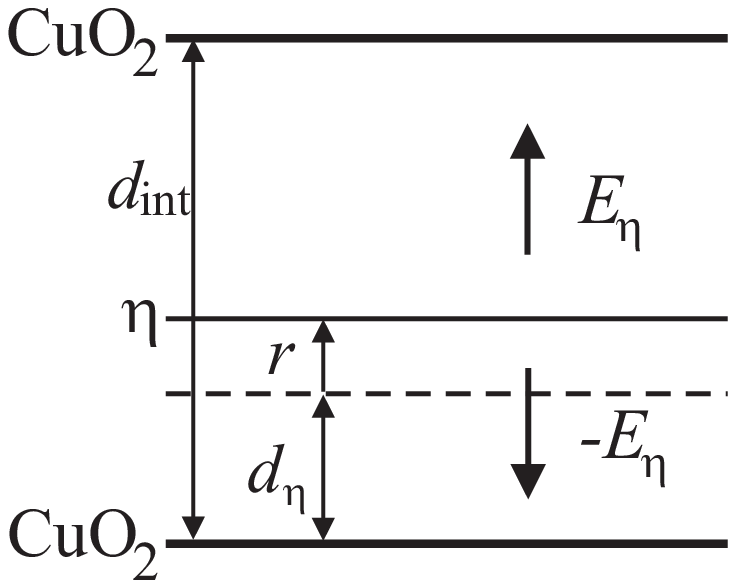}
  \end{center}

\newpage
Dubroka \etal\
Fig.~\ref{apict}\\[2cm]
  \begin{center}
  \includegraphics[width=14cm]{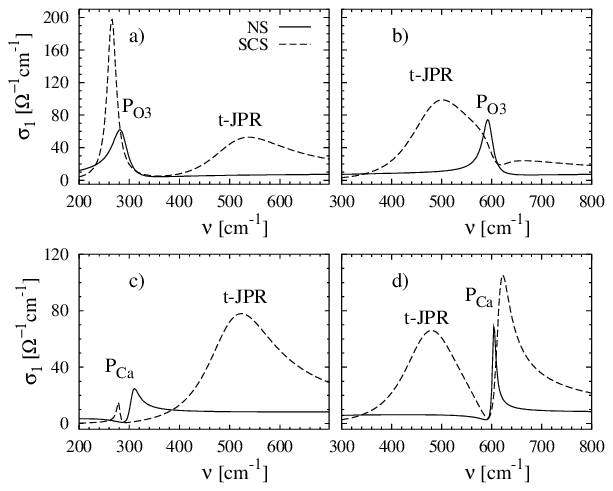}
  \end{center}

\newpage
Dubroka \etal\
Fig.~\ref{plane}\\[2cm]
  \begin{center}
  \includegraphics[width=14cm]{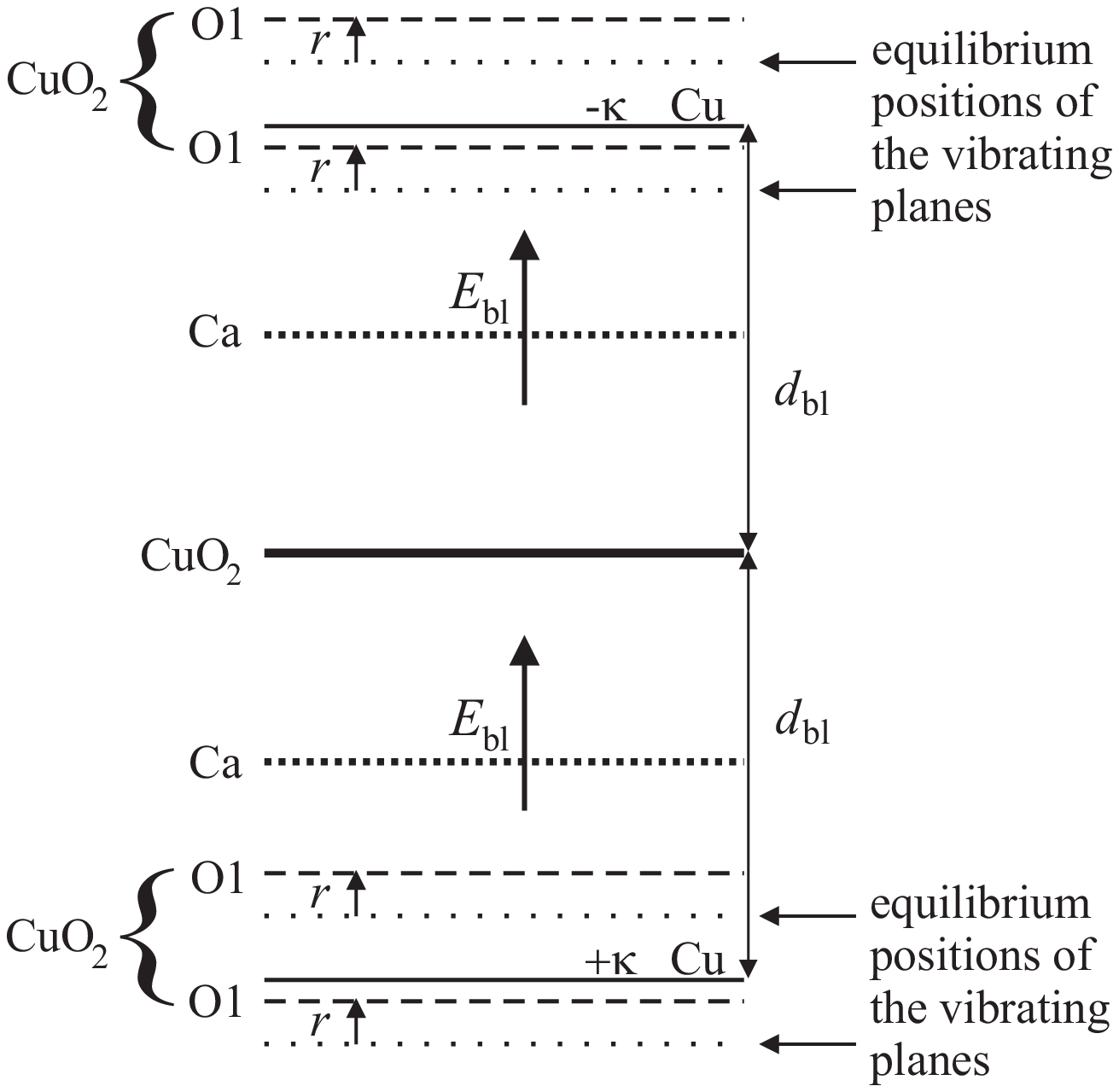}
  \end{center}

\newpage
Dubroka \etal\
Fig.~\ref{ocp}\\[2cm]
  \begin{center}
  \includegraphics[width=14cm]{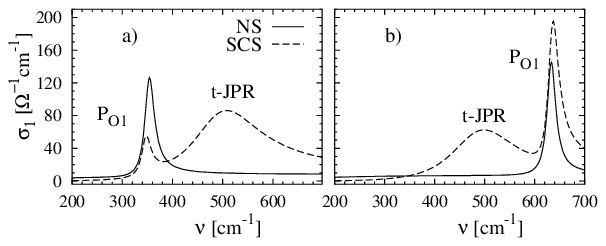}
  \end{center}

\newpage
Dubroka \etal\
Fig.~\ref{pruzina}\\[2cm]
  \begin{center}
  \includegraphics[width=7cm]{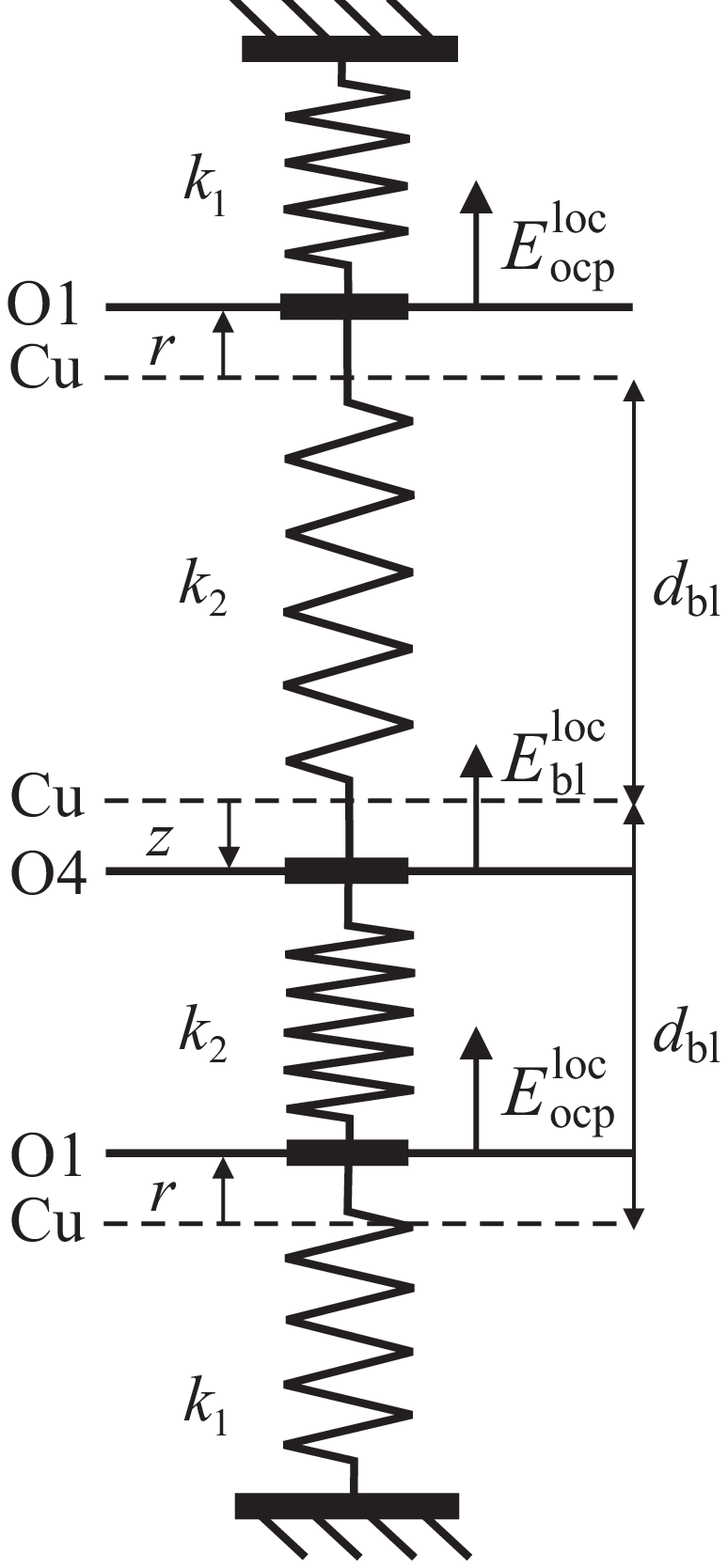}
  \end{center}

\newpage
Dubroka \etal\
Fig.~\ref{twin}\\[2cm]
  \begin{center}
  \includegraphics[width=14cm]{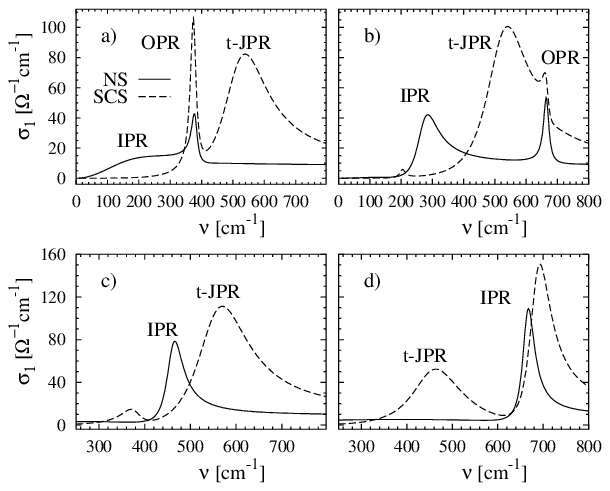}
  \end{center}

\newpage
Dubroka \etal\
Fig.~\ref{local2}\\[2cm]
  \begin{center}
  \includegraphics[width=15cm]{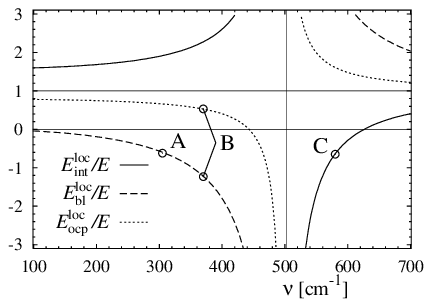}
  \end{center}

\newpage
Dubroka \etal\
Fig.~\ref{vodiv}\\[2cm]
  \begin{center}
  \includegraphics[height=18cm]{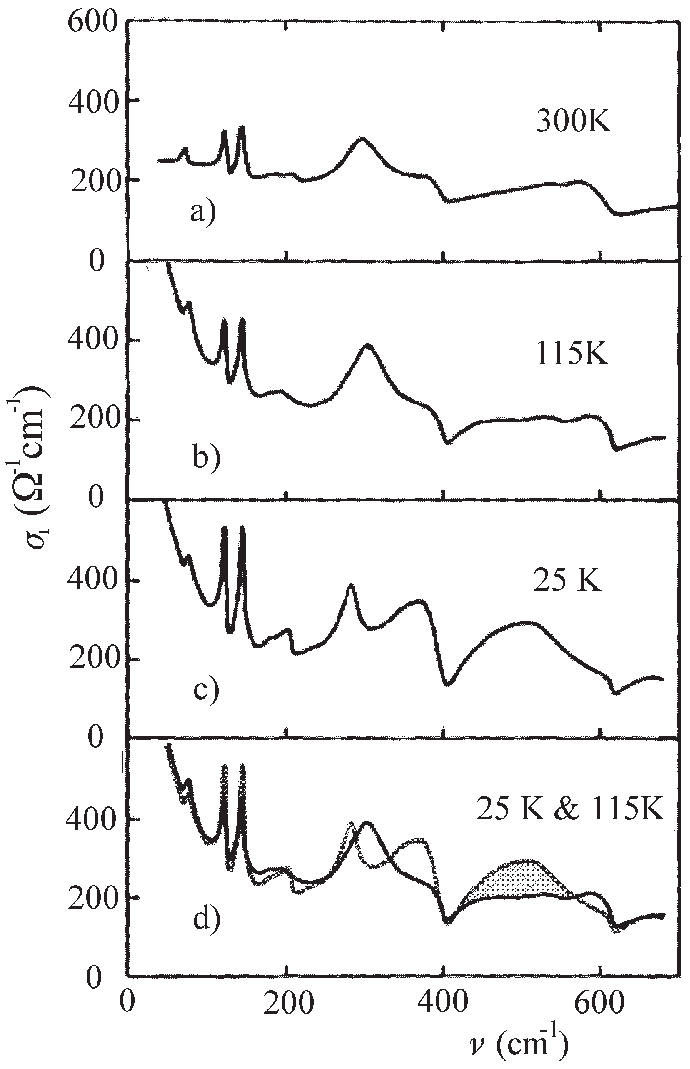}
  \end{center}

\newpage
Dubroka \etal\
Fig.~\ref{thtisk}\\[2cm]
  \begin{center}
  \includegraphics[width=14cm]{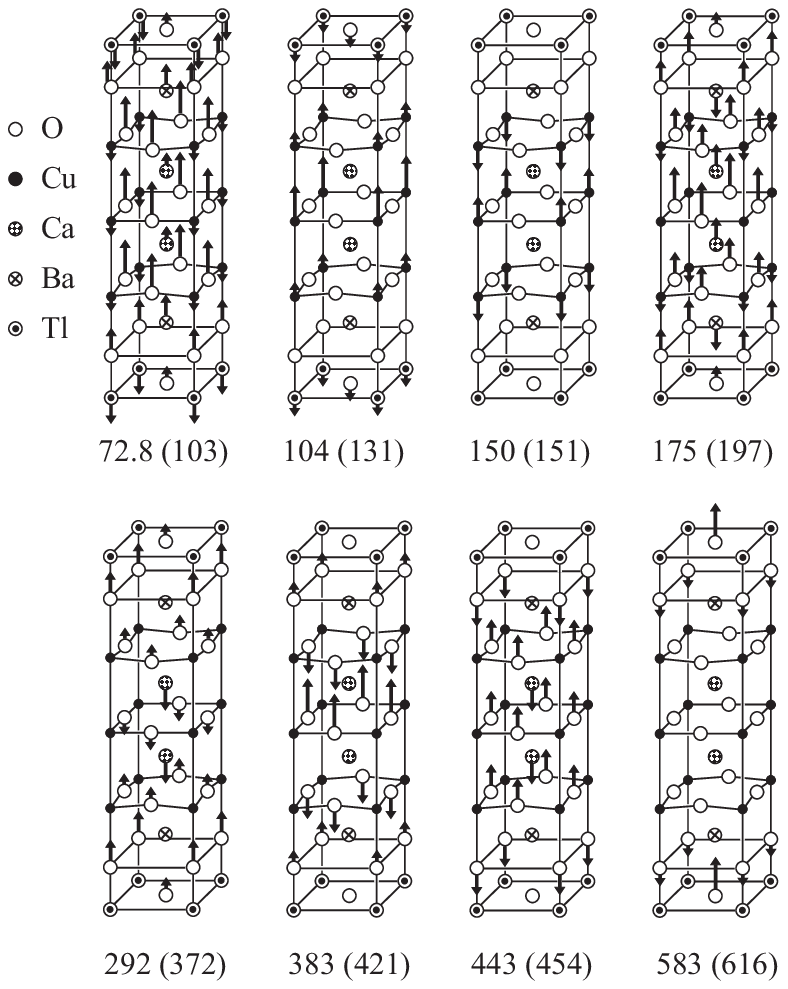}
  \end{center}


\newpage
\begin{thebibliography}{99}
\bibitem{Litvinchuk}{A.P.~Litvinchuk, C.~Thomsen and M.~Cardona,  {\sl
Physical Properties of High Temperature Superconductors IV}, editor
D.M.~Ginsberg Singapore World Scientific 1994.}
\bibitem{Homes}{C.C.~Homes, T.~Timusk, D.A.~Bonn, R.~Liang and
W.N.~Hardy, Physica C {\bf 254} (1995) 265.}
\bibitem{Schutzman}{J.~Sch\"utzmann, S.~Tajima, S.~Miyamoto, Y.~Sato and
R.~Hauff, Phys. Rev.~B {\bf 52} (1995) 13\,665.}
\bibitem{Bernhard}{C.~Bernhard, D.~Munzar, A.~Golnik, C.T.~Lin,
A.~Wittlin, J.~Huml\'\i\v{c}ek, M.~Cardona, Phys. Rev. B {\bf 61} (2000) 618.}
\bibitem{Zelezny}{V.~\v{Z}elezn\'y, S.~Tajima, D.~Munzar, T.~Motohashi,
J.~Shimoyama, and K.~Kishio, Phys. Rev. B {\bf 63} (2001) 060502.}
\bibitem{Boris}{A.V.~Boris, D.~Munzar, N.N.~Kovaleva, B.~Liang,
C.T.~Lin, A.~Dubroka, A.V.~Pimenov, T.~Holden, B.~Keimer,
Y.-L.~Mathis and C.~Bernhard, Phys. Rev. Lett. {\bf 89}
(2002) 277001.}
\bibitem{Zetterer}{T.~Zetterer, M.~Franz, J.~Sch\"utzman, W.~Ose,
H.H.~Otto, and K.F.~Renk, Phys. Rev.~B {\bf 41} (1990) 9\,499.}
\bibitem{Hadjiev}{V.G.~Hadjiev, Xingjiang Zhou, T.~Strohm, M.~Cardona,
Q.M.~Lin and C.W.~Chu, Phys. Rev. B {\bf 58} (1998) 1043.}
\bibitem{Limonov}{M.~Limonov, S.~Lee, S.~Tajima, A.~Yamanaka, Phys. Rev. B
{\bf 66} (2002) 054509.}
\bibitem{Petrov}{
J.H.~Chung, T.~Egami, R.J.~McQueeney, M.~Yethiraj, M.~Arai, T.~Yokoo,
Y.~Petrov, H.A.~Mook, Y.~Endoh, S.~Tajima, C.~Frost, F.~Dogan, Phys. Rev. B
67  (2003) 014517.}
\bibitem{Kulic}{M.L.~Kulic, Physics Reports {\bf 338} (2000) 1-264.}
\bibitem{Lanzara}{A.~Lanzara, P.V.~Bogdanov, X.J.~Zhou, S.A.~Kellar,
D.L.~Feng, E.D.~Lu, T.Yoshida, H.~Eisaki, A.~Fujimori, K.~Kishio,
J.I.~Shimoyama, T.~Noda, S.~Uchida, Z.~Hussain, Z.X. Shen, Nature {\bf 412}
(2001) 510.}
\bibitem{Shen}{Z.X.~Shen, A.~Lanzara, S.~Ishihara, N.~Nagaosa,
Philosophical Magazine B  {\bf 82} (2002) 1349.}
\bibitem{Mike}{D.~Munzar, C.~Bernhard, A.~Golnik, J.~Huml\'\i\v{c}ek,
M.~Cardona, Solid State Commun. {\bf 112} (1999) 365.}
\bibitem{Marel}{D.~van der Marel and A.~Tsvetkov,
Czechoslovak Journal of Physics {\bf 46} (1996) 3165.}
\bibitem{Gruninger}{M.~Gr\"uninger, D.~van der Marel, A.A.~Tsvetkov, and
A.~Erb, Phys. Rev. Lett. {\bf 84} (2000) 1575.}
\bibitem{Marel2}{D. van der Marel and A.A.~Tsvetkov,
Phys. Rev. B {\bf 64} 024530.}
\bibitem{Shah}{N.~Shah and A.J.~Millis, Phys. Rev. B {\bf 65} (2001)
024506.}
\bibitem{Ehrenreich}{H.~Ehrenreich, in {\it The Optical Properties of
Solids}, Proceedings of the International School of Physics ``Enrico
Fermi'', Course XXXIV, Varenna, 1965, edited by J.~Tauc (Academic Press,
New York, 1996), Chap. 13.}
\bibitem{Kittel}{Charles Kittel, {\sl Introduction to Solid State Physics}
(John Wiley \& Sons Inc., 1976, 5. edition), Chap. 13.}
\bibitem{Timusk}{T.~Timusk, C.C.~Homes, Solid State Commun. {\bf 126} (2003)
63.}
\bibitem{Hasegawa}{M.~Hasegawa, Y.~Matsushita, H.~Takei, Physica C {\bf
267} (1996) 31.}
\bibitem{Mike2}{D.~Munzar, C.~Bernhard, T.~Holden, A.~Golnik,
J.~Huml\'\i\v{c}ek and M.~Cardona, Phys. Rev. B {\bf 64} (2001) 024523.}
\bibitem{erratum}{D.~Munzar \etal, unpublished.}
\bibitem{Genzel}{L.~Genzel, A.~Wittlin, M.~Bauer, M.~Cardona,
E.~Sch\"{o}nherr, and A.~Simon, Phys. Rev. B {\bf 40} (1989) 2170.}
\bibitem{Kulkarni}{A.D.~Kulkarni, F.W.~de Wette, J.~Prade,
U.~Schr\"oder a W.~Kress, Phys. Rev.~B {\bf 41} (1990) 6\,409.}
\bibitem{Renk}K.F.~Renk,
{\it in Thallium-Based High-Temperature Superconductors},
editor A.M. Hermann, J.V.~Yakhmi, (Marcel Dekker Inc. New York  1994).
\bibitem{Renk2}{K.F.~Renk, W.~Ose, T.~Zetterer, J.~Schutzmann,
H.~Lengfellner, H.H.~Otto, J.~Keller, B.~Roas, L.~Schultz,
G.~Saemannischenko, Infrared physics {\bf 29} (1989) 791.}
\end{thebibliography}
\end{document}